\documentclass[a4paper,10pt]{article}

\usepackage[english]{babel}

\usepackage{amssymb,amsmath}
\usepackage{graphicx}
\usepackage{subcaption}
\usepackage{gensymb}
\graphicspath{{figs/}}
\usepackage[colorinlistoftodos]{todonotes}
\usepackage{xcolor}
\usepackage{xfrac}
\usepackage[normalem]{ulem}
\usepackage{tikz}
\usepackage{siunitx}
\usepackage{authblk}
\usepackage{cite}
\usepackage{tabularx}

\usepackage{hyperref}
\hypersetup{
	colorlinks,
	citecolor=blue,
	linkcolor=blue,
	urlcolor=blue
}

\newcommand{\av}[1]{\left\langle {#1} \right\rangle}
\newcommand{\vect}[1]{\mathbf{#1}}
\renewcommand{\vec}[1]{\mathbf{#1}}

\usepackage{setspace}
\usepackage[margin=2.5cm]{geometry}

\title{On local isotropy and scale dependence of pair dispersion in turbulent canopy flows}

\author[1]{Ron Shnapp}
\author[2]{Alex Liberzon}
\author[3]{Yardena Bohbot-Raviv}
\author[3]{Eyal Fattal}

\affil[1]{\small{Ben Gurion University of the Negev, Mechanical engineering department, POB 653, Beer Sheva 8410501, Israel}}
\affil[2]{\small{Tel Aviv University, School of Mechanical Engineering, POB 39040, Tel Aviv 6997801, Israel}}
\affil[3]{\small{Israel Institute for Biological Research, POB 19, Ness Ziona 7410001, Israel}}

\begin{document}

\onehalfspacing

\maketitle

\begin{abstract}
	Canopy flows in the atmospheric surface layer play important economic and ecological roles, governing the dispersion of passive scalars in the environment. The interaction of high-velocity fluid and large-scale surface-mounted obstacles in canopy flows produces drag and causes intense, inhomogeneous, and anisotropic turbulence. In this work, we focus on the turbulent dispersion of passive scalars by studying the ``pair dispersion'' - a statistical measure of relative motion between particles. We analyze the results of a 3D-PTV experiment in a wind tunnel canopy flow, focusing on small scales. We confirm the existence of local isotropy of pair dispersion at scales smaller than a characteristic shear length scale $L_\Gamma=(\epsilon/\Gamma^3)^{1/2}$, where $\epsilon$ and $\Gamma$ are the mean dissipation rate and shear rate, respectively. Furthermore, we show that pair dispersion in this locally isotropic regime is a scale-dependent super-diffusive process, similar to what occurs in homogeneous isotropic turbulent flows. In addition, we measure the pair relative velocity correlation function, showing that its de-correlation occurs in the locally isotropic regime, and discuss the implications of this observation for modeling pair dispersion. Thus, our study extends the fundamental understanding of turbulent pair dispersion to the anisotropic, inhomogeneous, turbulent canopy flow, bringing valuable information for modeling scalar dispersion in the atmospheric surface layer.
\end{abstract}

\section{Introduction}\label{sec:intro}

Transport, dispersion, and mixing are essential processes that determine a myriad of physical phenomena with crucial importance to the environment. These processes are driven by the random advection induced by turbulent flows. In particular, numerous such processes occur in the so-called canopy flows, namely in the lower part of the atmospheric surface layer. Examples include the dispersion of viral particles and fungal spores, the ventilation of urban air pollution, and the facilitation of vegetation transpiration by inducing humidity, CO$_2$ and heat fluxes. Our focus in this study is on pair dispersion in turbulent canopy flows, which involves studying the relative motion between Lagrangian fluid particles. As was shown by Batchelor~\cite{Batchelor1952}, the statistics of pair dispersion can be used to determine the variance of concentration fluctuations of advected passive scalars.

The study of canopy flows has garnered significant research attention over the past five decades, leading to a plethora of insightful discoveries. This field has witnessed substantial efforts, as evidenced by numerous comprehensive reviews that have effectively summarized the key findings~\cite{Raupach1981, Finnigan2000, Britter2003, Belcher2005, Brunet2020}. In neutrally stable conditions, canopy flows exhibit a distinctive feature characterized by the direct interaction between fluid and large-scale wall-mounted obstacles within a boundary layer. This interaction generates a drag force that decelerates the flow within the canopy layer ($z<H$, where $H$ represents the canopy height). At the upper boundary of the canopy ($z=H$), the drag discontinuity results in pronounced mean shear ($\partial U / \partial z$), giving rise to coherent Kelvin-Helmholtz structures akin to a mixing-layer analogy \cite{Raupach1996}. Turbulence statistics within the canopy layer are consequently both inhomogeneous and anisotropic. Eulerian single-point velocity probability density functions (PDFs) typically exhibit skewness, indicating intermittency and relatively high kurtosis values (Brunet, 1994). Interestingly, Lagrangian velocity increments have been observed to follow a Gaussian distribution \cite{Shnapp2020}. Additionally, turbulence in canopy flows is typically generated through two interdependent mechanisms: (I) shear production near $z \approx H$, and (II) production of smaller-scale turbulence within the wakes of the obstacles within the canopy ($z<H$) \cite{Finnigan2000}.

A common framework for modeling dispersion in canopy flows is through Lagrangian stochastic models~\cite{Poggi2006a, Aylor2001, Bailey2018, Baldocchi1997, DePaul1986, Raupach1987, Katul1997, Reynolds1998, Wilson2009, Duman2016, Fattal2021, Fattal2023}. In such models, Lagrangian fluid particles are advanced through the flow field using random walks that simulate the turbulent flow; this allows estimating statistics of the concentration field as a result of a certain distribution of scalar sources. Although these types of models allow us to bypass such difficulties as the hindering of parameterizations due to the existence of multiple scales or the inaccuracy of Taylor's frozen-turbulence hypothesis~\cite{Raupach1981, Raupach1987}, there are still numerous open issues in their development and formulation. For the case of single particle motion, these issues include the non-uniqueness of first-order Markov random walk models in three dimensions~\cite{Thomson1987}, the effects of coherent structures~\cite{Raupach1996, Ghisalberti2002}, non-Gaussian velocity PDFs~\cite{Pope1990}, the parallel contributions of wake and shear production~\cite{Shnapp2020}, or the mechanical diffusion~\cite{Nepf1999}. To the best of our knowledge, previous studies have only considered the motion of single particles, and there are no previous studies that deal with the relative motion between particles in canopy flows. Notably, a full description of the dispersion of a group of particles demands that the relative motion between any of their possible combinations be resolved~\cite{Batchelor1952}. Pair dispersion deals with the relative motion of pairs of particles, thus making a step forward in this respect.

In this work, we present an analysis of pair dispersion in a canopy flow using results from a wind tunnel experiment. This flow mimics the neutrally buoyant atmospheric surface layer in an environmental wind-tunnel setup, as we reported in Ref.~\cite{Shnapp2019}. We focused in the past on single particle statistics of this canopy flow, revealed short decorrelation Lagrangian timescales~\cite{Shnapp2020}, and characterized the unique features of Lagrangian intermittency in this flow~\cite{Shnapp2021}. Our results deal with relatively small scales and are relevant for describing dispersion in between roughness elements inside the canopy. These findings are useful for instance for modeling dispersion between buildings in urban areas. 

Background on pair dispersion and details on the experiment are given next. Following that, our analysis is divided into two main parts -- first we quantify the effect of the mean shear and anisotropy in our canopy model through the pair dispersion tensor. Following that, we study the different regimes of pair dispersion that are observed in our measurement, including the ballistic and the inertial regime with a scale-dependent diffusivity; we suggest how these results could be useful for estimating the concentration variance of released passive scalars in the flow. Thus, our results could be useful for constructing and validating turbulent dispersion models in the atmospheric surface layer.

\subsection{Background on pair dispersion in isotropic and anisotropic turbulence}

Turbulent pair dispersion was introduced by Richardson~\cite{Richardson1926} and since then it has been studied extensively, e.g. as reviewed in Refs.~\cite{Salazar2009, Falkovich2001, Sawford2001}. Pair dispersion describes the statistics of vector $\vec{r}$ connecting two, initially close, Lagrangian fluid particles:
\begin{equation}
\vec{r} = \vec{x}^{(1)} - \vec{x}^{(2)} \qquad ; \qquad r \equiv |\vec{r}| \, ,
\label{eq:r}
\end{equation}
where $\vec{x}^{(i)}$ is the 3D coordinate of a particle $i$ (bold symbols denote vectors). Leaning on principles of homogeneous isotropic turbulence (HIT), the dynamics of $r$ can be divided into three distinct regimes. (I) At short times, there is a ballistic regime~\cite{Batchelor1952}, during which particles' relative motion is co-linear. (II) At larger time scales, Richardson~\cite{Richardson1926} suggested that statistics of $r$ can be modeled as a diffusive process with a scale-dependent diffusivity parameter, $K$:
\begin{equation}
K \equiv \frac{1}{2} \frac{\partial \av{r^2}}{\partial t} \qquad \text{with} \qquad K \propto \langle r^2 \rangle^{2/3} \sim r^{4/3}
\label{eq:k_scaling}
\end{equation}
\noindent where $\langle \cdot \rangle $ is an ensemble average of many particle pairs. The scaling of $K$, commonly termed ``the four-thirds law'', leads to a super-diffusive growth of the separation distance with  $\av{r^2} \sim \tau^{3}$~\cite{Monin1972}. (III) at later times when the two particles have separated farther away than the integral turbulence scale, i.e. for $r \gg L$, the distance $r$ is expected to be diffusive with a constant, or Taylor, diffusivity~\cite{Taylor1921}:
\begin{equation}
K = T_L \tilde{u}^2
\label{eq:taylor_diffusivity}
\end{equation}
\noindent where $T_L$ is the Lagrangian timescale and $\tilde{u}$ is the root-mean-square (RMS) of velocity fluctuations.
Therefore, considering an ensemble of pairs of particles, initially separated by $r_0$ and moving in a HIT flow, the second order moment of the distance, the variance of the change in separation distance is summarized as follows~\cite{Salazar2009, Monin1972}:
\begin{equation}
\av{(r-r_0)^2} = \begin{cases}
\frac{11}{3}C_2 (\epsilon r_0)^{2/3} \tau^2  \quad \quad  &   \text{for} \,\,\,\, \tau \ll \tau_b \\
g \epsilon \tau^3  &  \text{for} \,\,\,\,  \tau_b \ll \tau \ll T_L \\
2 \tilde{u} T_L \tau  &  \text{for} \,\,\,\,  T_L \ll \tau
\end{cases}
\label{eq:separation_laws}
\end{equation}
where $C_2\approx2.1$~\cite{Sreenivasan1995} is the so-called Kolmogorov constant, $\epsilon$ is the mean rate of the turbulent kinetic energy dissipation, $g\approx0.5$~\cite{Ott2000} is a so-called universal Richardson constant, and the Batchelor timescale is defined as:
\begin{equation}
\tau_b = \left( \frac{r_0^2}{\epsilon} \right)^{1/3} \, .
\label{eq:tb}
\end{equation}

The theory of pair dispersion has been critically revised and extended in recent years. In particular, the three regimes have been tested in several experiments and numerical simulations for various initial distances, $r_0/\eta$  (where $\eta=(\nu^3/\epsilon)^{1/4}$ is the Kolmogorov dissipation length scale).
Verification tests of Eq.~\eqref{eq:separation_laws} in Refs.~\cite{Yeung1994, Ott2000, Boffetta:2002, Biferale2005, Bourgoin2006, Ouellette2006a, Berg2006} have shown that it is difficult to observe the transition from ballistic to the super-diffusive regime, partly because it requires long-duration tracking, i.e. for $\tau\gg\tau_b$, and high Reynolds number flows~\cite{Bourgoin2006, Ouellette2006, Berg2006}. Specifically, the super-diffusive regime is asymptotic, and so, finite Reynolds number effects cause the evolution of statistics of $r$ to depend strongly on the initial separation, $r_0/\eta$~\cite{Biferale2005}. Another issue is the strong intermittency of pair dispersion~\cite{Biferale2014,  Bitane2012, Scatamacchia2012, Thalabard2014, Shnapp_Liberzon:2018}, presumably due to Lagrangian statistics associated with long-time correlations~\cite{Boffetta2002}. In particular, recent observations have shown that the scaling of $\av{r^2}$ depends on the initial conditions, namely on the initial separation~\cite{Elsinga2021, Tan2022} and the initial separation velocity~\cite{Shnapp_Liberzon:2018}, while it is also known that particles may retain their separation distance for very long times~\cite{Scatamacchia2012}. 
Thus, it was suggested that Richardson's diffusive approach relies on an assumption of short Lagrangian correlation times~\cite{Bitane2012}, which could explain the inconsistency with the recent observations. Several models using different forms of the so-called persistent ballistic random walks have been proposed as alternatives~\cite{Sokolov2000, Rast2011, Thalabard2014, Bourgoin2015}. Bitane et al.~\cite{Bitane2012} proposed an alternative to $\tau_b$ which renders the process self-similar with respect to $r_0$; this framework is discussed in more detail below.

While Eq.~\eqref{eq:separation_laws} and the studies mentioned above pertain to HIT flows, pair dispersion in inhomogeneous or anisotropic flows has received much less attention. In inhomogeneous and anisotropic flows, statistics vary for each component of $\vec{r}$, and their evolution may be co-dependent on each other. In particular, in anisotropic flows it is necessary to introduce a pair-dispersion tensor with components~\cite{Batchelor1952}:
\begin{equation}
\Delta_{ij}(\tau) = \av{ 
	\big(r_i - r_{i,0} \big) \cdot \big(r_j - r_{j,0}\big)} \,\, ;
\label{eq:PD_Tensor}
\end{equation}
notably, $\mathrm{tr}(\Delta_{ij}) = \av{(r-r_0)^2}$, and in HIT $\Delta_{ij}$ is diagonal. Furthermore, pair dispersion in anisotropic flows can depend on the initial orientation of $\vec{r}_0$ with respect to the production mechanisms of turbulent kinetic energy; for shear flows, such as the canopy flow considered here, this would be the mean shear direction.

Shear effects on pair dispersion were studied previously in homogeneous shear flows through computational simulations. Shen and Yeung~\cite{Shen1997} established that the presence of a mean shear has a strong effect on pair dispersion: it makes the evolution of $\Delta_{ij}$ anisotropic and with non-zero non-diagonal components, and it can render apparent super-diffusivity at long times ($\Delta_{xx}\sim \tau^3$) irrespective of the argument of scale-dependence. More recently, Pitton et al.~\cite{Pitton2012} and Polanco et al.~\cite{Polanco2018} studied pair dispersion in turbulent channel flows and highlighted the combined effects of both anisotropy and inhomogeneity with respect to the distance from the wall. Their analysis clearly showed that where the mean shear is strongest the anisotropy of pair dispersion is most dominant. In a DNS of a square duct flow, Sharma and Phares~\cite{Sharma2006} reported that particles remained trapped in secondary flow regions which affected the power law behavior of $r(\tau)$. Celani et al.~\cite{Celani2005} have proposed that mean shear and turbulent fluctuations act on pair dispersion on different time scales, which leads to two distinct regimes of either turbulent fluctuations dominance or mean shear dominance. A crossover between these two regimes was predicted to occur at a critical timescale that is proportional to the mean shear $\tau_c\sim \left(\frac{dU}{dz} \right)^{-1}$.  Recent studies further examined buoyancy-driven flows~\cite{Schumacher2008, Ni2013, Liot2019} and pair dispersion of inertial particles~\cite{Pitton2012} or in other complex flows, such as in the stroke of a swimming jellyfish~\cite{Kim2019}.

Despite the extensive body of work, the above discussion establishes the existence of a wide gap in the understanding of turbulent pair dispersion. Indeed, even for ideal cases, such as HIT or homogeneously sheared turbulence, our understanding is lacking and the introduction of flow inhomogeneity brings on another dimension of complexity. In this study, we bring the first empirical results on pair dispersion from an inhomogeneous and anisotropic canopy flow in order to shed some light on the prevailing issues.


\section{Methods}

\subsection{A 3D particle-tracking wind-tunnel experiment}

We study pair dispersion in a canopy flow using the results of a 3D particle tracking velocimetry experiment (3D-PTV~\cite{Dracos1996}). In the experiment, we obtained flow tracer trajectories in a wind-tunnel canopy flow model using an extended, real-time image processing system~\cite{Shnapp2019}. The wind tunnel laboratory, situated at the Israel Institute for Biological Research (Ness Ziona, Israel), features a 14 meters long open circuit suction wind tunnel with a $2\times2$ squared meters cross-sectional area, that is fit for conducting experiments that mimic turbulent flows in the atmospheric surface layer~\cite{Bohbot-Raviv2017}. The canopy flow was modeled by placing flat rectangular plates along the entire bottom wall of the test section. 
Our study used an inhomogeneous canopy layer, constructed with two types of flat plates with heights either $H$ or $\frac{1}{2}H$, and width $\frac{1}{2}H$, where $H=100 \si{\milli \meter}$; the two types of plates were positioned in consecutive rows and at a staggered construction, see Fig~\ref{fig:locs}. 
The canopy frontal area density, defined as $\Lambda_f = A_f /A_T$, ($A_f$ being the element frontal area, and $A_T$ the lot area of the canopy) is $\frac{1}{2}$, which positions our canopy right between the ``dense" and ``sparse" categories~\cite{Brunet2020}. 
Data was gathered at two levels of the free stream velocity, $U_\infty = 2.5$ and $4 \si{\meter \per \second}$ being the free stream mean velocity measured with a sonic anemometer at the center of the wind tunnels' cross-section. These values correspond to Reynolds numbers $Re_\infty \equiv \frac{U_\infty H}{\nu} \approx 16\times10^3$ and $26\times10^3$, where $\nu$ is the kinematic viscosity of the air.

Our previous work with this canopy construction, based on particle image velocimetry (PIV) measurements, showed that the double-averaged velocity profile in our canopy layer had a smaller shear length at the top of the canopy as compared to homogeneous canopy layers at similar densities and that, unlike homogeneous canopies, it had a second inflection point right above the lower roughness elements~\cite{Shig2023}. Furthermore, as commonly observed in canopy flows, sweep events were observed inside the canopy layer ($z<1.5H$), albeit with a second peak of increased sweep contribution above the lower set of elements~\cite{Shig2023a}. In addition, our previous analysis of 3D-PTV measurements showed that the canopy is characterized by an unexpectedly short Lagrangian de-correlation timescale, leading to probability density functions of velocity increments that are of a Gaussian shape~\cite{Shnapp2020} and complemented by intermittency, as observed in single particle Lagrangian statistics~\cite{Shnapp2021}.

In the experiment, we tracked the positions of flow tracers using the 3D-PTV method through an extended 3D-PTV application. The details of the methods are given in Ref.~\cite{Shnapp2019}, so only brief information shall be repeated here for completeness. The flow was seeded with hollow glass spheres with an estimated Stokes number of $\overline{St}\approx 0.05$, indicating good tracking properties with only a minor filtration of high-frequency contents of the turbulent dynamics that cannot be ruled out. The tracer particles were illuminated with a 10W continuous wave laser beam as shown in Fig~\ref{fig:locs_a}. Using a set of four 4-megapixel cameras, we obtained 3D particle trajectories at a resolution of 50$\si{\micro \meter}$ per pixel, inside volumes of measurement were roughly $8\times5\times4$mm$^3$ in size. We implemented camera calibration, stereo matching, and tracking~\cite{Dracos1996}, to reconstruct the tracer particle's trajectories by using the openPTV~\cite{openptv} open-source software, integrated to operate with the real-time image analysis extension. The data analysis was performed using the Flowtracks package~\cite{Meller2016}. In this work, we use the frame of reference that is commonly used in the canopy flow literature: $x$ is aligned with the streamwise direction longitudinally within the wind tunnel, $y$ is the cross-stream horizontal direction, and $z$ points vertically up against gravity and away from the bottom wall.

\subsection{Data analysis: a quasi-homogeneous approach}

\begin{figure}[h!]
	\centering
	\begin{subfigure}[t]{0.47\textwidth}
		\centering
		\includegraphics[height=5.5cm]{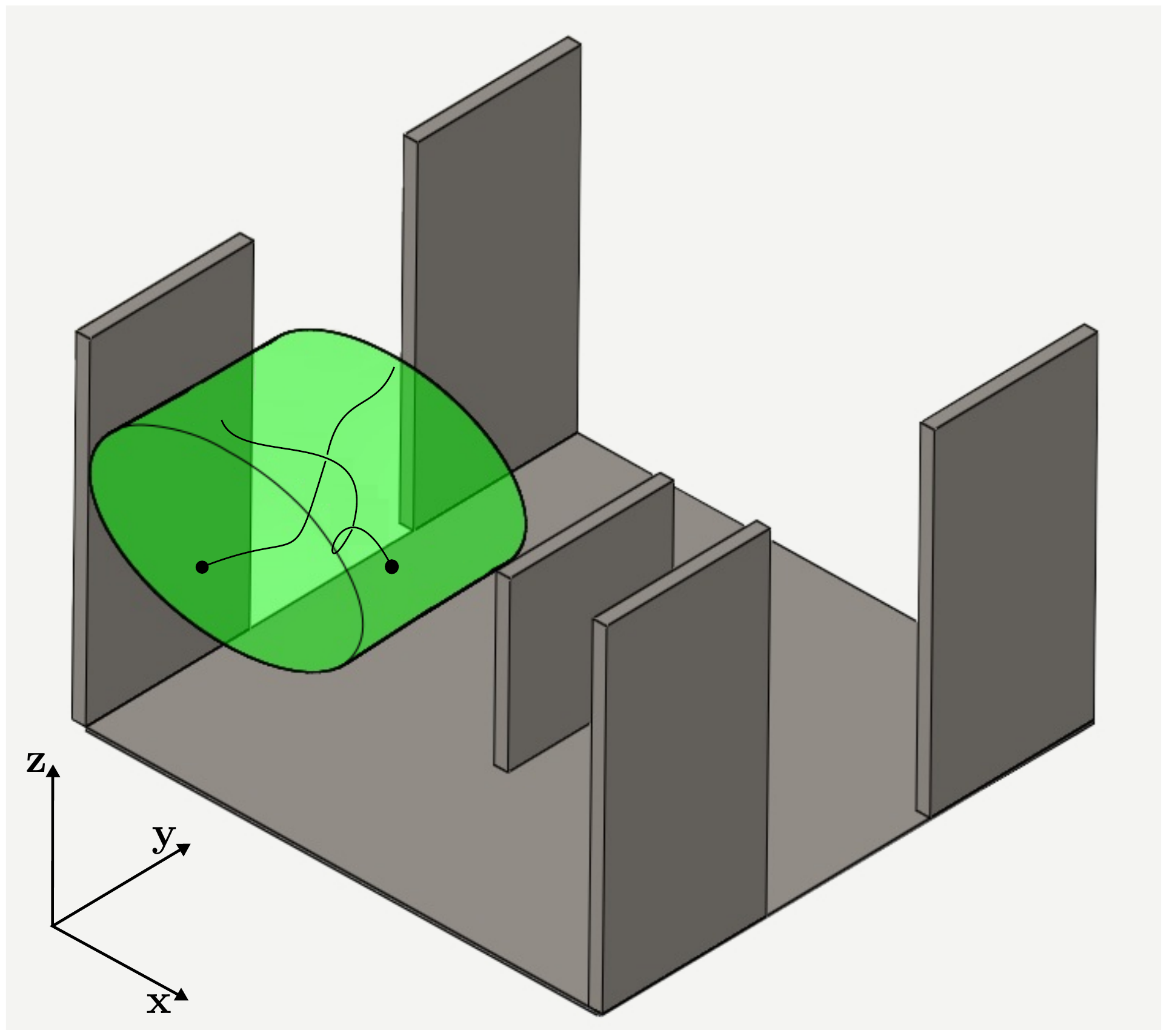} 
		\caption{\label{fig:locs_a}}
	\end{subfigure}
	\begin{subfigure}[t]{0.47\textwidth}
		\centering
		\includegraphics[height=5.5cm]{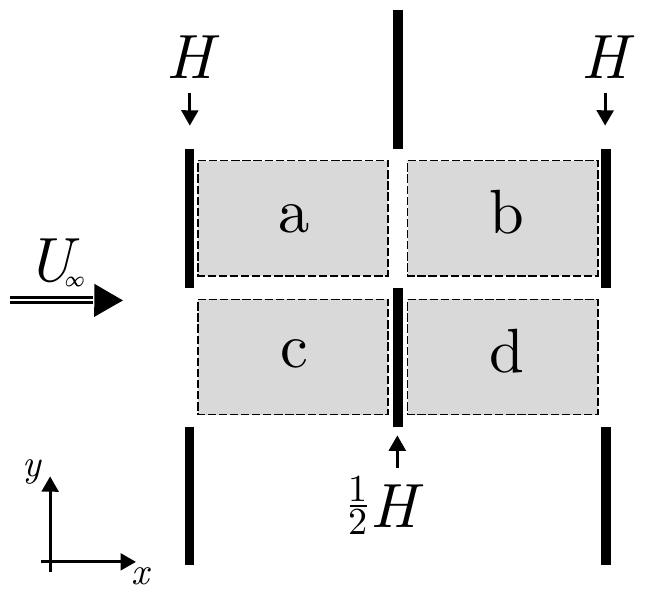}
		\caption{\label{fig:locs_b}}
	\end{subfigure}
	
	\caption{Sketches representing the area of measurement. (a) an isometric view of the coordinate system, several roughness elements, and a cutout laser beam illuminating a single sub-volume with tracer particles. (b) a top view of the measurement volume showing the 4 horizontal sub-volume positions. \label{fig:locs}}
\end{figure}

We recorded trajectories in the height range $0.5H < z < 1.5H$ and across a single representative canopy unit cell, by scanning the full volume through 20 sub-volumes, as depicted by the green shaded region in Fig.~\ref{fig:locs_a}. The sub-volumes were defined at 4 horizontal locations, and at 5 different heights above the wind-tunnel bottom wall. Sub-volume horizontal positions are labeled alphabetically $a$, $b$, $c$ and $d$ as shown in Fig.~\ref{fig:locs}(b). At each position, $a-d$, we used 5 vertical slabs of thickness $0.2 H$ to define the sub-volumes; the vertical slab position is labeled numerically 1-5 and corresponds to heights $z/H \in$ \{0.5-0.7, 0.7-0.9, 0.9-1.1, 1.1-1.3, 1.3-1.5\}, respectively. The volume scanning approach allowed us to measure a full canopy unit cell, however, it limited the particle trajectory lengths to be within each of the individual sub-volumes.

In the frame of our analysis, we present statistical properties calculated over ensembles of trajectories. Thus, formally, the mean value of any Lagrangian quantity $A$ is estimated using the  average -
\begin{equation}
\av{A} \equiv \frac{1}{N}\sum\limits_{i=0}^N A_i(\vec{x}_{\mathbb{V}},t),
\label{eq:mean}
\end{equation} 
\noindent where $\mathbb{V}$ is a tag pertaining to different particle ensembles. In our recent work~\cite{Shnapp2020}, we explored single-particle statistics using the same dataset. There, the sub-volume averaged quantities approach was proposed, in which Lagrangian statistics were presented for ensembles of particles found at each sub-volume, essentially resulting in spatially averaged statistics across each small sub-volume. It was furthermore shown that this quasi-homogeneous approach is justified in our flow because random turbulent forcing out-weighted combined effects of flow inhomogeneity terms in the stochastic particle equation of motion. Specifically, this was achieved through the condition:
\begin{equation}
\av{ \frac{1}{2} \frac{|C_0 \epsilon R_{ij}^{-1} u_j'|}{|\phi_i /g|} } \ll 1
\label{eq:quasi-homogeneity}
\end{equation}
\noindent where $C_0$ is the so-called Lagrangian structure function coefficient, 
$R_{ij}=\av{u'_i \, u_j'}$ is the Reynolds stress tensor, and $\phi_i/g$ is a vector representing a sum of drift terms of the Lagrangian velocity PDF that allows accounting for flow inhomogeneity~\cite{Thomson1987}. In the present work, we capitalize on the condition~\eqref{eq:quasi-homogeneity}, and present quasi-homogeneous statistics for pair dispersion as well. This point is important since it validates the effective scaling laws we obtain below. 

In addition to that, in Ref.~\cite{Shnapp2020} turbulent parameters of our flow were calculated where they were discussed and analyzed in depth. In particular, we calculated $\epsilon$, the turbulence dissipation length and time scales $\eta$, $\tau_\eta$, the Lagrangian decorrelation timescale $T_{i}$ and the Taylor Reynolds number $Re_\lambda$. The values of these parameters are tabulated for each sub-volume in Appendix~\ref{app1}, and they shall be used in the analysis below.

%
%

\section{Local isotropy of pair dispersion}\label{sec:local_isotropy}

\subsection{Scaling analysis}\label{sec:scaling}

As discussed in Sec.~\ref{sec:intro}, numerical simulations~\cite{Shen1997, Sharma2006, Pitton2012, Polanco2018, Polanco2019} confirmed theoretical predictions~\cite{Monin1972} that the mean shear, inhomogeneity, and/or anisotropy in the flow cause anisotropic pair dispersion. It is based on the fact that the fundamental mechanism driving pair separation is the relative velocity between Lagrangian particles. Because turbulent flow statistics are scale-dependent, it is expected that also the "degree" of anisotropy in pair dispersion will depend on the separation distance $r$. Turbulent velocity fluctuations are characterized by a tendency to recover isotropy at small scales due to the isotropy of the equation of motion. The return to isotropy at smaller scales was also observed in canopy flows~\cite{Stiperski2021}. This suggests that at sufficiently small values of $r$, and as long as the Reynolds number is sufficiently high, pair dispersion will be, at least approximately, isotropic, as predicted theoretically in Ref.~\cite{Celani2005}. Thus, to determine whether local isotropy or anisotropy is expected to occur in our pair dispersion measurement, we first conduct an analysis of the relevant scales of the flow.

Anisotropic turbulent fluctuations are introduced to the flow by the boundary conditions. In the canopy flow case, the flow impinges on the roughness obstacles that exert drag and retard the flow. Thus, the production of turbulence in canopies is commonly decomposed into two main contributions~\cite{Finnigan2000}: I) shear production at the top of the canopy layer, and II) production at smaller scales in the wakes of the obstacles. The third possible component, the production by wall friction is usually negligible compared to form drag due to the low velocity near the ground for sufficiently dense canopies. Because shear production relates to the horizontally averaged mean shear, it induces an intrinsic anisotropy in the structure of turbulence in canopy flow. On the other hand, wake production occurs due to local variations in the rates of strain, associated with the wakes and boundary layers of individual roughness element~\cite{Finnigan2000}.
Since the local orientation of most straining directions changes in space, we expect that the anisotropy in pair dispersion statistics will mostly reflect shear production, at least to a first approximation. This seems to be particularly true for Lagrangian statistics such as in pair dispersion since the particles essentially sample different flow regions.

\begin{figure}
	\centering
	\includegraphics[width=1.0\textwidth]{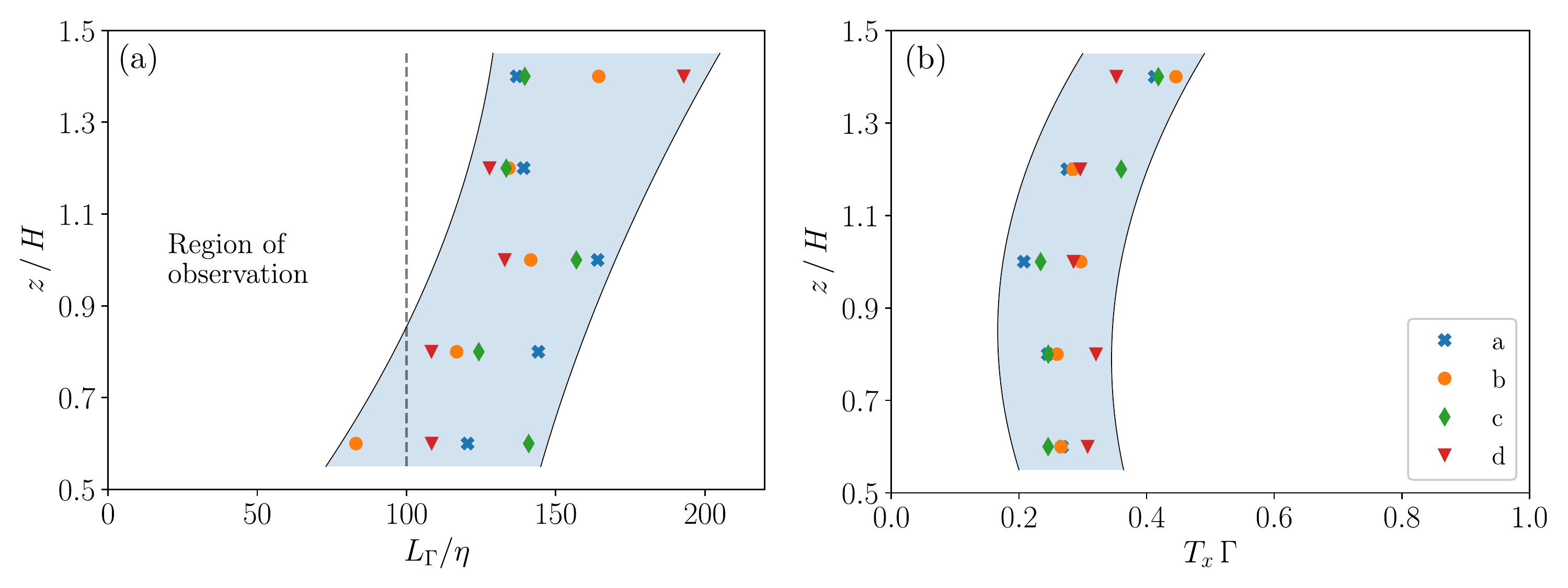}
	\caption{(a) Ratio of the shear length scale and the Kolmogorov length scale for the various sub-volumes plotted as a function of height. (b) The Lagrangian velocity decorrelation timescale, $T_x$, normalized using the mean shear rate plotted as a function of height. Data is shown for the $Re_\infty=26\times10^3$ case. \label{fig:scaling_analysis}}
\end{figure}

The appropriate length scale that characterizes the production of turbulence by shear is
\begin{equation}
L_\Gamma = \left( \frac{\epsilon}{\Gamma^3} \right)^{1/2}
\label{eq:shear_length}
\end{equation}
where $\Gamma = \frac{dU}{dz}$ is the mean velocity gradient. Indeed, analyzing the scale-by-scale turbulent kinetic energy budget in turbulent shear flows, Casciola et al.~\cite{Casciola2003} showed that the production by shear continues down to $L_\Gamma$, whereas for scales $r<L_\Gamma$ the TKE transport becomes more dominant. For our analysis, we parameterize the shear length using a global, canopy-wide measure, of the mean velocity gradient - $\Gamma_z = \frac{U(z=1.5H)}{1.5H} \approx 4.75s^{-1} $. Figure~\ref{fig:scaling_analysis}(a) shows that $L_\Gamma/\eta$ increased from roughly 100 inside the canopy and up to roughly 150 above it. 

Another critical dimension of pair dispersion is time. Celani et al.~\cite{Celani2005} predicted that for particles with sufficiently small initial separations, pair dispersion is initially isotropic while it becomes anisotropic once the particles have grown sufficiently far apart for the anisotropic turbulent scales to be prominent. They further noted that the critical timescale for this transition is proportional to the inverse mean shear rate, $\tau_c \sim \Gamma^{-1}$. Figure~\ref{fig:scaling_analysis}(b) shows the Lagrangian velocity decorrelation timescale for the streamwise velocity component normalized by $\Gamma$. Throughout our measurement region, we observe that $T_x \cdot \Gamma<1$.

Due to the finite measurement region, our work focuses on timescales up to $\tau \approx T_x$. In addition, the typical separations between particles that we consider in our work reach up to roughly $r=100\eta$, namely mostly values with $r<L_\Gamma$. Thus, the dimensions of our measurement region and the timescales on which we focus are smaller than the scales imposed by the mean shear. For this reason, we expect to see only weak anisotropy in pair dispersion statistics due to the local isotropy of the flow.

\subsection{Observation of weak anisotropy}\label{sec:weak_anisotopy}

\begin{figure}[th!]
	\centering
	\includegraphics[width=0.75\textwidth]{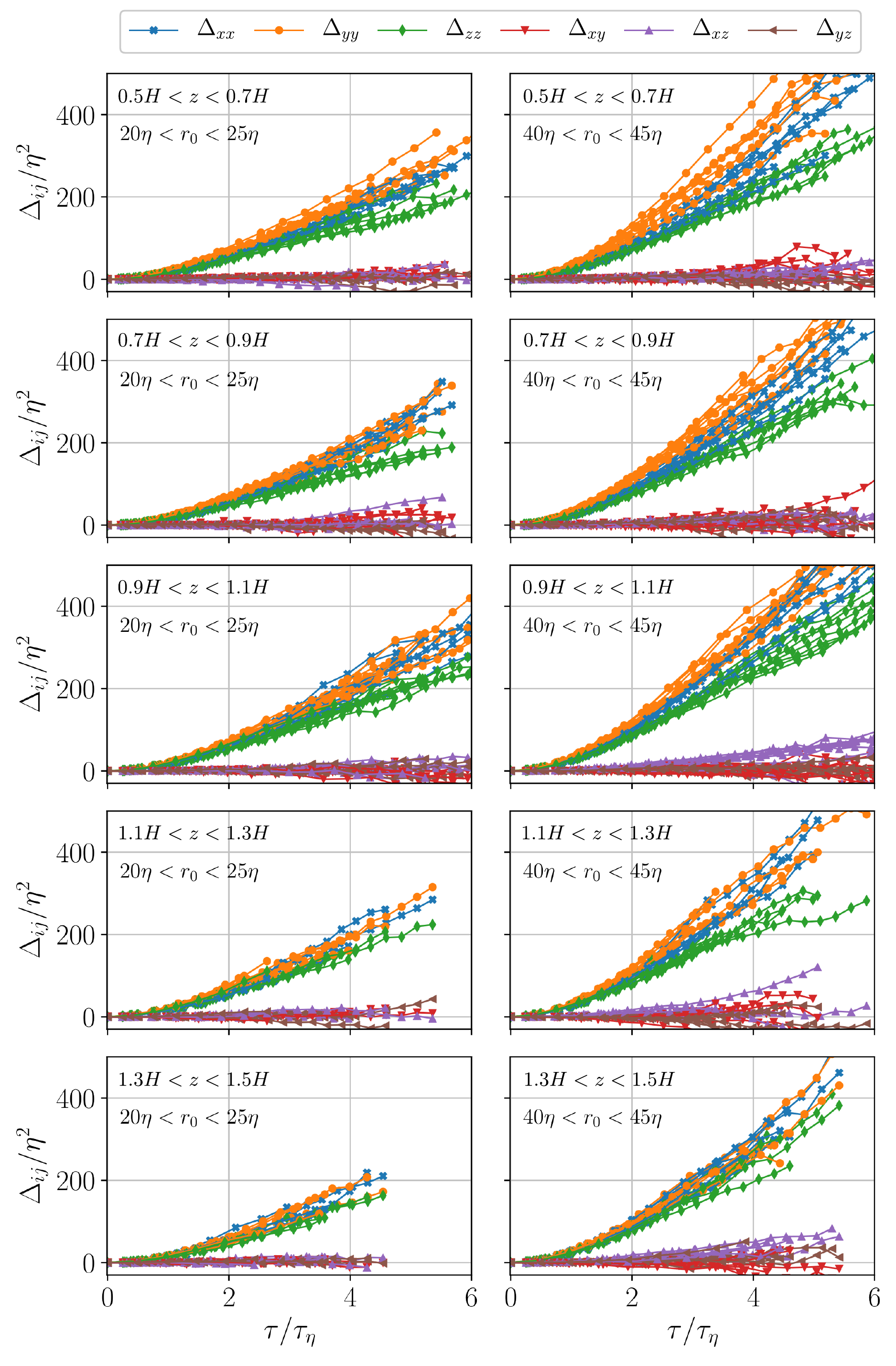}
	\caption{The various components of the pair dispersion tensor are shown for two initial separation values and at the five height groups used. Different shapes correspond to different components of $\Delta_{ij}$. Lines with the same shape come from each of the four horizontal sub-volume locations and thus represent the horizontal variability of the statistics. \label{fig:Dij_components}}
\end{figure}
The various components of the pair dispersion tensor, $\Delta_{ij}$, across the whole region of measurements are shown in Fig.~\ref{fig:Dij_components} for two representative initial separation values inside the inertial range. In all cases, the diagonal components steadily increase from zero as time increases, which indicates increasing separations in all orthogonal directions as expected. The off-diagonal components, on the other hand, generally do not increase significantly in the available time frame of the measurements, with a trend for larger scatter for the larger initial separation cases and at longer times. This occurs consistently across the entire measurement sub-volumes.

Since the non-diagonal components represent mixed averages of the separation vector's components, their relatively low values indicate a weak correlation among the separation in different directions. This behavior is unlike what is expected in a mean shear-driven separation scenario as, for example, in a homogeneously sheared turbulence the separation velocity in the streamwise direction grows as the separation along the mean velocity gradient direction increases. Thus, the behavior observed for $\Delta_{ij}$ is consistent with the notion presented in Sec.~\ref{sec:scaling} that the mean shear is not expected to significantly affect pair dispersion in our measurements.

The three diagonal components of the pair dispersion tensor grow with time, and slightly different values are seen for the various components. To highlight these differences we introduce the following tensor 
\begin{equation}
I_{ij} \equiv \frac{\Delta_{ij}}{\mathrm{tr}(\Delta_{ij})}-\frac{1}{3}\delta_{ij} 
\label{eq:anisotropy}
\end{equation}
where $\delta_{ij}$ is the identity matrix.
The tensor $I_{ij}$ shows statistics of the components of $\vect{\Delta r}$ with respect to its norm, essentially highlighting anisotropy. Let us note that $\mathrm{tr}(I_{ij})=\sum\limits_i I_{ii} = 0$, and that dispersion in the direction of components with $I_{ii}>0$ is faster than in the isotropic case (i.e., than $\frac{1}{3}\av{(r-r_0)^2}$), while components with $I_{ii}<0$ the dispersion is slower. In a limiting case in which pairs separate only in one direction, say $x$, then $I_{xx}=\frac{2}{3}$ while $I_{yy}=I_{zz}=-\frac{1}{3}$.

\begin{figure}[h]
	\centering
	\includegraphics[width=.75\textwidth]{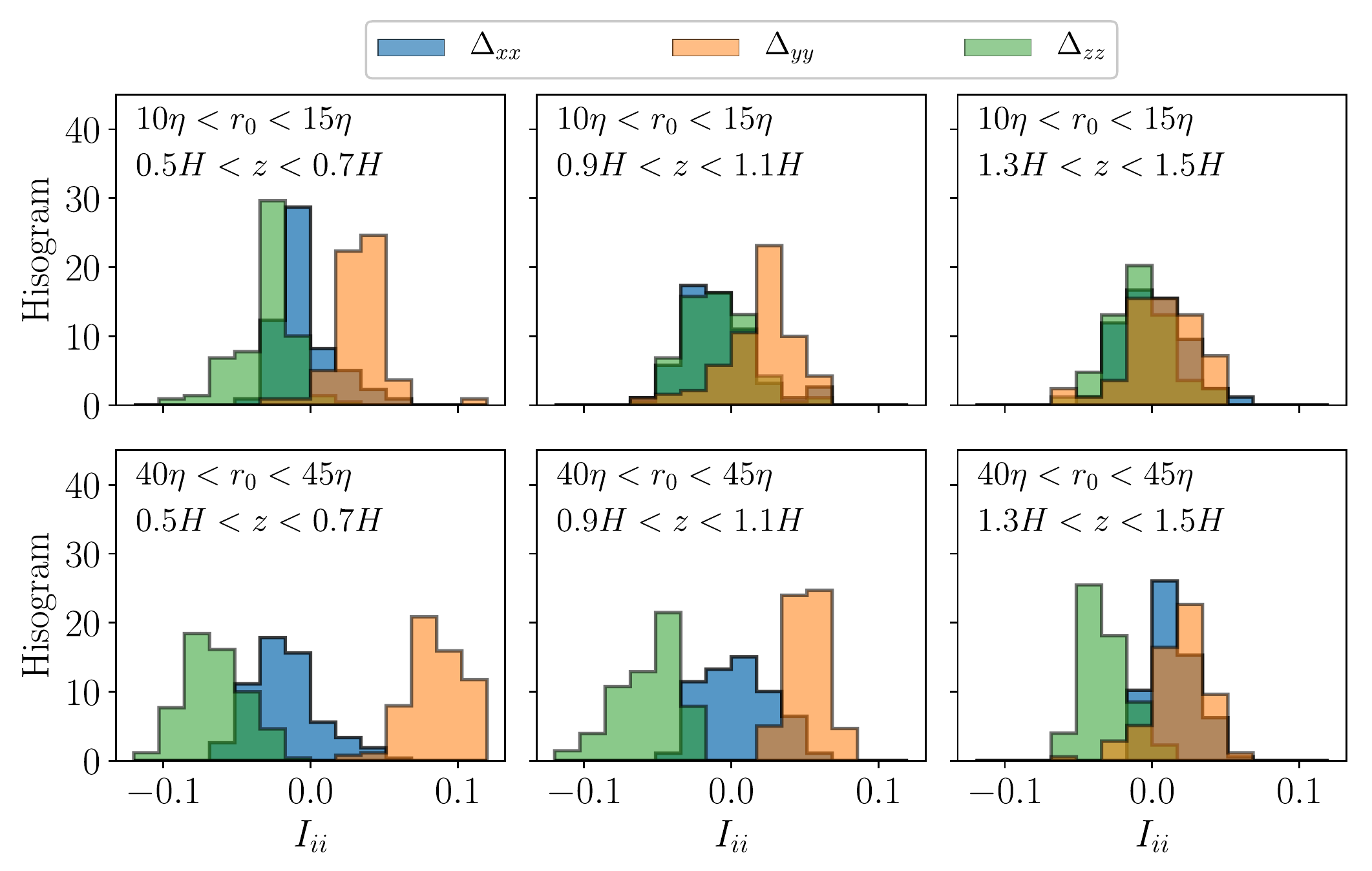}
	\caption{Diagonal terms of the pair dispersion tensor normalized by its trace minus one-third. Data is shown for sub-volumes at three heights and for two levels of $r_0$. \label{fig:PD_Anisotropic}}
\end{figure}

\begin{figure}[h]
	\centering
	\includegraphics[width=.8\textwidth]{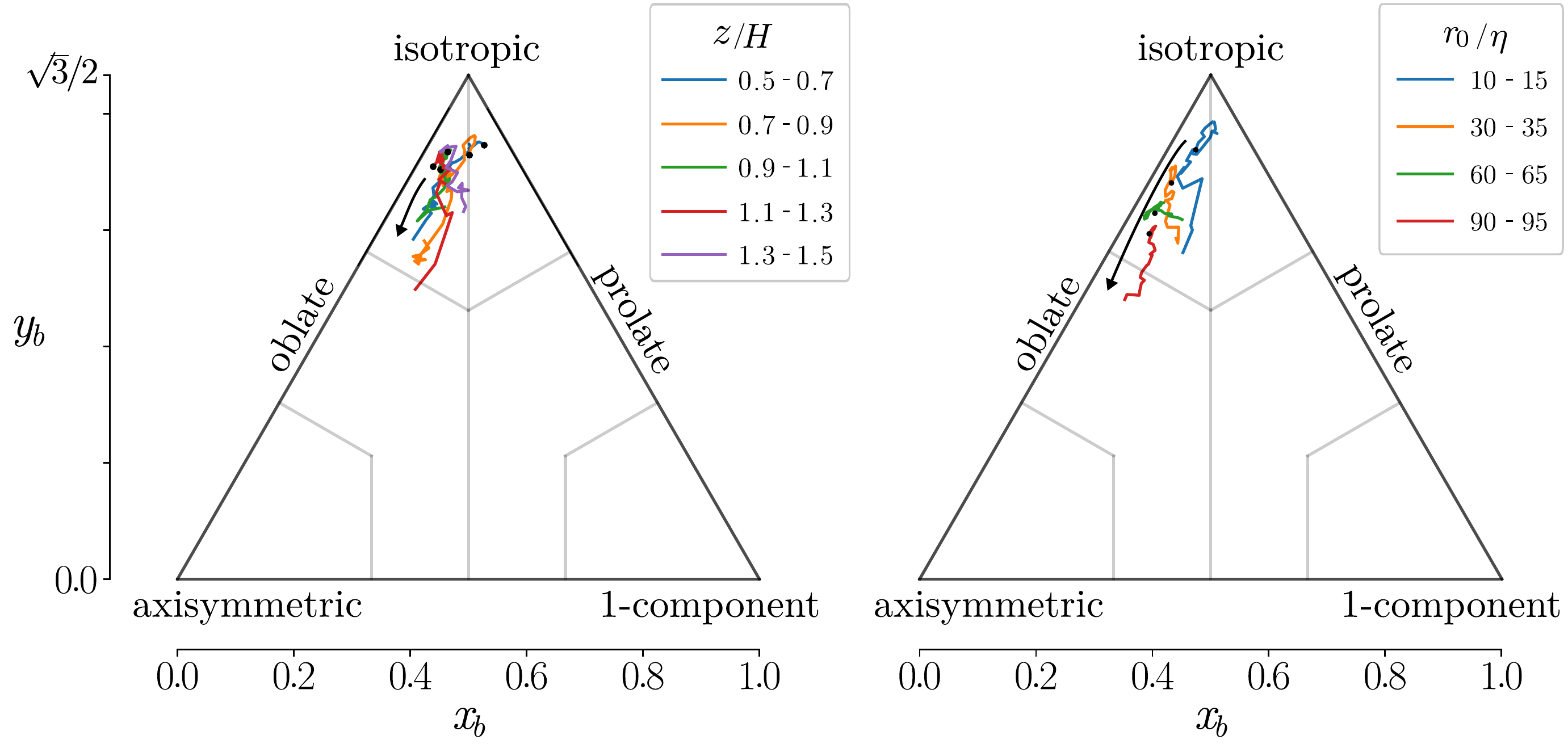}
	\caption{Trajectories of pair dispersion anisotropy on the $x_b$ $y_b$ plane. Two datasets are shown: for five heights with $15\eta r_0<20\eta$ (left) and for four initial separation values with $0.9H<z<1.1H$ (right). The beginning of each trajectory, i.e. at time zero, is marked by a black circle, where the trajectories from which the trajectories evolve with time up to 7$\tau_\eta$. Data is for the $\mathrm{Re}_\infty=2.6\times10^4$ case, and horizontally averaged across all sub-volumes. \label{fig:anisotropy_triangle}}
\end{figure}

Histograms of the diagonal components, $I_{ii}(\tau)$ are shown in Fig.~\ref{fig:PD_Anisotropic}, focusing on two representative initial separations and three representative heights: inside the canopy $0.5 < z/H < 0.7$, at its top $0.9 < z/H < 1.1$, and above it $1.3 < z/H < 1.5$. The histograms count data points from all horizontal positions, both $Re_\infty$, and across the time range $\tau<6\tau_\eta$, during which $I_{ii}$ did not change appreciably.

The disparity between the separation components decreases with height, being the largest inside the canopy and weakest above it. Furthermore, the separation was the slowest in the vertical direction and fastest in the spanwise direction, whereas the streamwise separation is approximately at isotropic values ($I_{xx}\approx0$). The fact that the separation is fastest in the spanwise direction and inside the canopy, suggests that the observed anisotropy could be attributed to the channeling of the flow in between the roughness obstacles, which were in a staggered configuration, and not necessarily due to the effects of the mean vertical shear. This would highlight the effects of dispersive fluxes between the canopy obstacles in driving horizontal dispersion. And still, we note that the magnitudes of $I_{ii}$ values reach only up to approximately 0.1; this corresponds to a weak anisotropy with 20\% of disparity between the separation in the vertical and spanwise components. In particular, this degree of anisotropy is much smaller than what was observed in previous works concerning flows with mean shear \cite{Shen1997, Pitton2012, Polanco2018}.

The disparity between the components
is also seen to increase with the initial separation, $r_0$. This is consistent with the picture of return to isotropy, as anisotropy becomes more prominent at larger scales~\cite{Lumley1977}.

The increasing of anisotropy with scale brings up questions regarding the development of anisotropy in pair dispersion with time. As particles separate, the growth of typical $r$ values suggests that pairs are exposed to the more anisotropic turbulent scales as time increases. The development of anisotropy could be examined by adopting a framework analogous to the one introduced by Stiperski et al.~\cite{Stiperski2021}, which used a projection of the invariants of the normalized Reynolds stress tensor on a two-dimensional plane, allowing them to investigate trajectories of the return to isotropy in various canopy flows. In an analogy to that, we examine here the projection of the eigenvalues of $I_{ij}$ using the same projection:
\begin{equation}
\begin{split}
x_b &= \lambda_1 - \lambda_2 + \frac{1}{2}\,(3\lambda_3 + 1)\\
y_b &= \frac{\sqrt{3}}{2} \, (3\lambda_3 + 1)
\end{split}
\label{eq:xb_yb}
\end{equation} 
where $\lambda_1$, $\lambda_2$, and $\lambda_3$, are the smallest intermediate and largest eigenvalues of $I_{ij}$ respectively. Equation~\ref{eq:xb_yb} maps the eigenvalues to a planar equilateral triangle. As seen in Fig.~\ref{fig:anisotropy_triangle}, the triangle's nodes correspond to cases of fully isotropic, 2-component axisymmetric, or one-component dispersion. Plotting the trajectories that correspond to the measured $I_{ij}$ tensor allows to probe the topology of pair dispersion as it varies in time. For almost all of the data shown, the weak anisotropy of pair dispersion is seen by the fact that the trajectories are almost completely confined to the isotropic part of the map. Nevertheless, the weak anisotropy that does exist is evident in the general trend of trajectories to progress along the left flank of the triangular maps, in the direction corresponding to the oblate topology. The trajectories also demonstrate that most of the anisotropy in our pair dispersion measurements is explained by the larger initial separation values and not due to the increase in separation with time. This is in line with the fact that our measurements are confined to time scales smaller than $\Gamma^{-1}$.

\subsection{Bias due to initial orientation}\label{sec:initial_orientation}

As shown in Refs.~\cite{Shen1997, Pitton2012, Polanco2018}, in flows with mean shear, pair dispersion is affected by the initial orientation of the pairs for short times. Indeed, taking as an example the case of a shear flow where the velocity is $\vect{u}= (\Gamma \, z, \, 0, \, 0)$ and considering particles with an initial separation $\vect{r}_0=(r_{0,x}, \, r_{0,y}, \, r_{0,z})$, then the separation vector   is $\vect{r}(\tau) = (r_{0,x} + \tau \, \Gamma \, r_{0,z}, \, r_{0,y}, \, r_{0,z})$. Therefore, even in the simplest scenario, the presence of mean shear biases the streamwise separation component. The bias depends on the projection of the initial separation vector on the mean shear direction, so for pairs whose $\vect{r}_o$ is aligned with the mean shear direction, the bias is by a factor of $\Gamma r_{0}$.

The relative importance of this bias in a turbulent flow can be estimated by comparing $\Gamma r_0$ with the spatial fluctuating velocity increments $\delta_{r_0} u$. If we apply the Kolmogorov dimensional scaling~\cite{Kolmogorov1941} to parameterize $\delta_{r_0} u \sim (\epsilon r_0)^{1/3}$, we can construct a dimensionless group $S = \frac{\Gamma\,r_0}{\delta_{r_0} u} = \frac{\Gamma r_0^{2/3}}{\epsilon^{1/3}}= \left( \frac{r_0}{L_\Gamma} \right)^{2/3}$ that quantifies the importance of the shear bias for particles with $r$ in the inertial range. When $S \gg 1 $ we expect that shear will bias pairs separation based on their initial orientation whereas when $S \ll 1 $ it will not. Also, if an inertial range exists, this bias grows stronger as $r_0^{2/3}$ for smaller initial separations, which is in qualitative agreement with the observations of Refs~\cite{Shen1997, Polanco2018}. Also, in the dissipation range, the velocity field is presumably smooth and the typical size of velocity gradients is $\frac{1}{\tau_\eta} = (\frac{\epsilon}{\nu})^{1/2}$ (where $\tau_\eta$ is the dissipation time scale). Therefore, the dissipation scaling for velocity increments in the dissipation range is $\delta_{r_0} u \sim  r_0 \left( \frac{\epsilon}{\nu} \right)^{1/2}$ so that the dimensionless parameter becomes $S = \Gamma \left( \frac{\epsilon}{\nu} \right)^{-1/2}$. To summarize, we define 
\begin{equation}
S = 
\begin{cases}
\Gamma \left( \frac{\epsilon}{\nu} \right)^{-1/2} & \quad \text{if} \qquad r_0\ll \eta\\
\left( \frac{r_0}{L_\Gamma} \right)^{2/3} & \quad \text{if} \qquad r_0\gg \eta
\end{cases}
\end{equation}  
\noindent and when $S\gg1$ the presence of mean shear is expected to bias pair dispersion, making it faster when the initial orientation of particles is aligned with the shear direction.

\begin{figure}[h]
	\centering
	\includegraphics[width=0.8\textwidth]{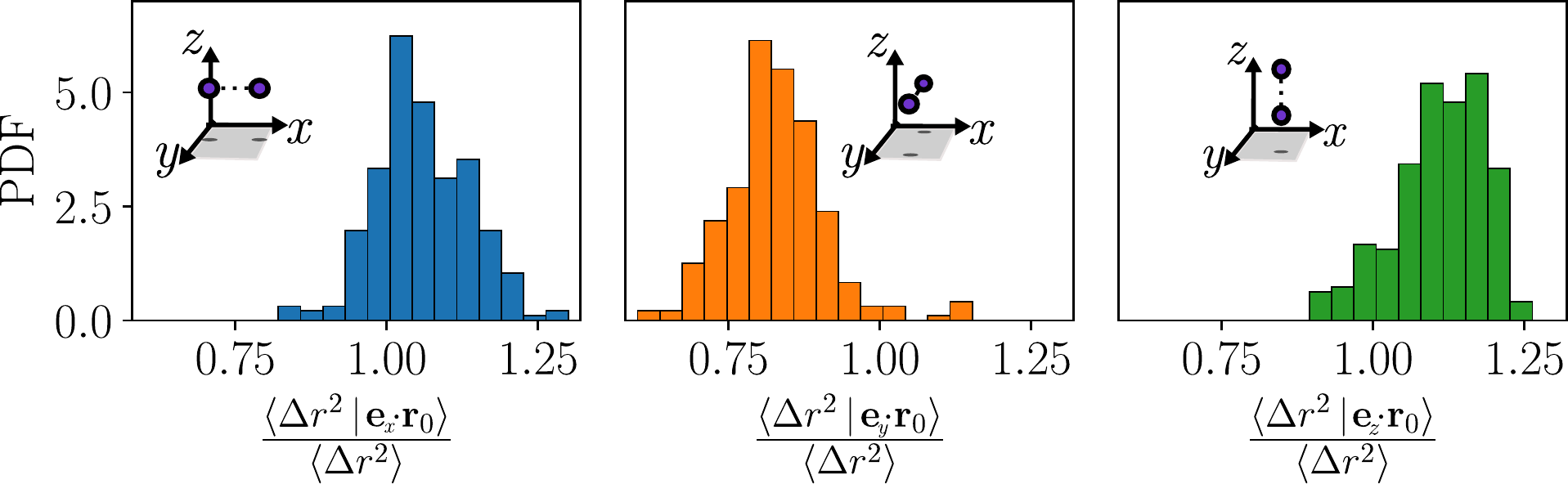}
	\caption{Probability distributions of the variance of change in separation distance conditioned on the initial orientation and divided by the unconditioned value. Data is taken at different times and all sub-volumes for pairs with $30\eta < r_0 < 50\eta$. \label{fig:init_ori}}
\end{figure}

To examine the effect of initial orientation in the canopy flow, we estimated the variance $\av{\delta r^2}$ conditioned on the initial orientation of $\vec{r}_0$. We divided the pairs into sub-samples based on the condition that the angle between $\vec{r}_0$ and each of the coordinate axes was lower than $25\degree$.
In Fig.~\ref{fig:init_ori}, probability distributions are shown for the conditional variance normalized by the variance of all the pairs. In this case, data was taken at all available times, all sub-volumes, and with $30\eta < r_0 < 50 \eta$. The distributions show that pairs with initial separation aligned vertically (i.e. with $\hat{z}$) typically separated faster than the average, whereas pairs with initial separation aligned with the spanwise (i.e. $\hat{y}$) typically separated slower than the average. Considering the previous observation that separations are fastest in the spanwise direction (Fig.~\ref{fig:PD_Anisotropic}), the bias observed here for vertically oriented pairs may suggest that the channeling effect varies with height. This is also consistent with the observation of the strongest anisotropy inside the canopy layer.

From Fig.~\ref{fig:scaling_analysis} it is seen that for the present case, we have a maximum value of $S\approx0.38$ which suggests that the bias should be quite weak. This is validated in Fig.~\ref{fig:init_ori} since it shows that the magnitudes of the pair dispersion bias encountered in our measurements reached up to 30\% and was typically around 15\%. These values are indeed rather low considering that previous studies~\cite{Pitton2012, Polanco2018} observed orders of magnitudes of difference between different groups of conditioned particles in regions of the flow with strong shear.

\medskip

Overall, our empirical results agree well with the scaling argument presented above and are consistent with the theory of Celani et. al~\cite{Celani2005}. These results reveal the important role of $L_\Gamma$ and $\Gamma$ in determining the scales relevant for the development of pair dispersion anisotropy in canopy flows. Indeed, anisotropy was weak at the scales relevant to our work, in addition to a weak bias due to the initial orientation of pairs. These key results highlight local isotropic turbulent fluctuations as the main driver of particle separations in the canopy flow at small scales.


\section{Regimes of pair dispersion and a scale dependent diffusivity}\label{sec:scale_dependent_diffusivity}

Following the observation in Sec.~\ref{sec:local_isotropy} that pair dispersion is only weakly anisotropic in our measurements we focus next on statistics of the change of the separation distance, $r=|\vec{r}|$, with time. In particular, we analyze the scaling of the change in separation distance $\av{(r-r_0)^2} \propto \tau^\beta$~\cite{Biferale2005, Berg2006, Ouellette2006a, Salazar2009, Shnapp_Liberzon:2018}. It is also noted that, formally, due to the inhomogeneity of our canopy flow, $\beta$ is essentially an "effective scaling". Therefore, the goal of this section is to compare the empirical data with the different regimes of Eq.~\eqref{eq:separation_laws}, in order to understand the phenomenology underlying pair dispersion in our canopy flow.

In order to investigate time scaling, it is essential to consider a possible bias due to the finite volume of measurement. In particle tracking experiments there is a concern that finite volumes could lead to a bias in the estimate of Lagrangian scaling exponents since particles leave the volume at different times. A similar bias when estimating Lagrangian correlation functions was investigated in Ref.~\cite{Biferale2008}. Specifically, the extreme cases in which particles separate very fast and leave the volume of measurement at short times could influence the slopes of the curves leading to errors in estimating the scaling exponent $\beta$. To avoid such a bias, the data were sub-sampled for pairs tracked for a fixed amount of time. For our analysis, we consider only pairs that were successfully tracked for durations $t_{tr} > 5 \tau_\eta$. Statistics were calculated only for $\tau \leq 5 \tau_\eta$ (where $t_{tr}$ is the tracking time, the duration of time for a particle to remain in the sub volume). This sub-sampling may cause an underestimation of the rate of separation. However, it is necessary for making reliable estimations of the scaling exponent. In addition to that, to maintain a concise description we present data for only one sub-volume -- b3, albeit the results were observed to be robust for particles from the other sub-volumes as well.

\subsection{The ballistic regime}

We first verify that pair dispersion at short times follows a ballistic propagation, as was suggested by Batchelor~\cite{Batchelor1952}. This regime is characterized by a quadratic growth of $\av{\Delta r^2}\propto \tau^2$ ($\tau \ll \tau_b$), corresponding to the first case of Eq.~\eqref{eq:separation_laws}. Fig.~\ref{fig:PD_b3_longtime} (a) presents pair dispersion from the canopy flow in normalized form according to Eq.~\eqref{eq:separation_laws},
\begin{equation}
\frac{\av{\Delta r^2}}{S_{LL}(r_0) \,\, \tau_b} = \left( \frac{\tau}{\tau_b} \right)^2 \,\, ,
\label{eq:ballistic_normalized}
\end{equation}
\noindent where $S_{LL}(r_0) = \av{ (\delta_{r_0} \vect{u} \cdot \frac{\vect{r_0}}{r_0})^2}$ is the directly measured Eulerian structure-function, and the Batchelor timescale $\tau_b$. Initially, for $\tau < 0.1 \tau_b$, all the curves with different initial separations collapse on a quadratic dashed line, as Eq.~\eqref{eq:separation_laws} suggests. Similar results were previously obtained in homogeneous flows by Refs.~\cite{Ouellette2006a, Bitane2012, Shnapp_Liberzon:2018}, and inhomogeneous flows by Refs.~\cite{Polanco2018, Liot2019, Ni2013}. The collapse of the data at short times using the longitudinal structure function $S_{LL}(r_0)$ is important since it confirms that $\left[S_{LL}(r)\right]^{1/2}$ is the appropriate velocity scale for the separation of particles at a distance $r$.

\subsection{Transition from a ballistic to an inertial regime}

\begin{figure}[t!]
	\centering
	\includegraphics[height=5.5cm]{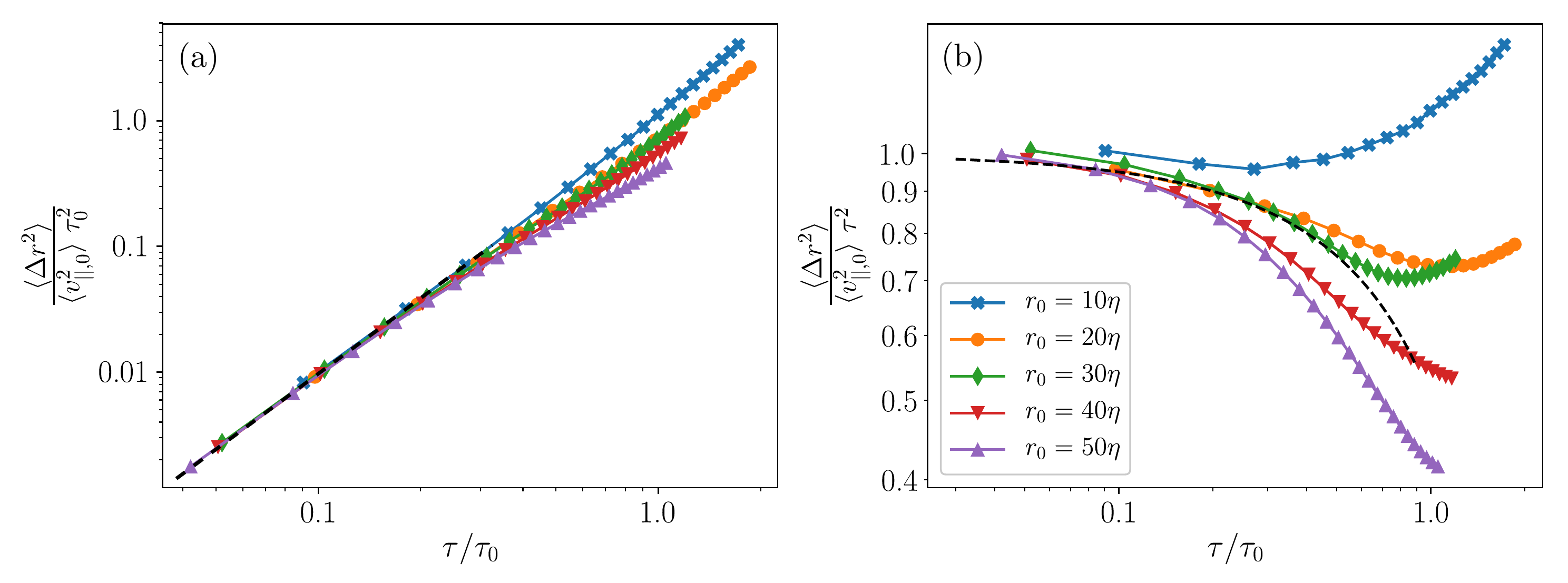}
	\caption{(a) Variance of the change in pair separation normalized with the second moment of separation velocities and the Batchelor timescale. The dashed line corresponds to the Ballistic regime, Eq.~\eqref{eq:ballistic_normalized}. (b) Same data but normalized using the time lag, and the dashed line corresponds to Eq.~\eqref{eq:ballistic_expanded}.
		\label{fig:PD_b3_longtime}}
\end{figure}

Next, we examine the termination of the ballistic regime, which is done following the analysis performed recently by Bitane et al.~\cite{Bitane2012} for the case of HIT flow. We denote the separation velocity $v_{||} = \frac{\partial r}{\partial t}$ and acceleration $a_{||} = \frac{\partial v_{||}}{\partial t}$. Then, the separation distance is Taylor expanded to the third order, squared, and ensemble-averaged, which gives
\begin{equation}
\av{\Delta r^2} = \av{v_{||,0}^2} \tau^2 + \frac{1}{2} \av{a_{||,0} \, v_{||,0}} \tau^3 + \mathcal{O}(\tau^4)
\label{eq:separation_Taylor_expansion}
\end{equation}
where subscript $0$ denotes initial values at $\tau=0$. This expression, which extends the Ballistic regime, is expected to hold for short times, and, in particular, according to Bitane et al.~\cite{Bitane2012}, it gives the appropriate timescale for the end of the ballistic regime at $ \tau_0 = -\frac{\av{v_{||,0}^2} }{ \av{a_{||,0} \, v_{||,0}}}$. Note that the minus sign is used since in 3D turbulence $\av{a_{||,0} \, v_{||,0}} < 0$~\cite{Ott2000}, as was confirmed for our data (not shown for brevity). Therefore, the above Taylor expansion can be rewritten in dimensionless form as 
\begin{equation}
\frac{\av{\Delta r^2}}{\big\langle v_{||,0}^2 \big\rangle \, \tau^2} = 1 - \frac{1}{2} \left( \frac{\tau}{\tau_0} \right) + \mathcal{O}(\tau^2)\,\, .
\label{eq:ballistic_expanded}
\end{equation}
In Fig.~\ref{fig:PD_b3_longtime}(b) we show the empirical data of $\av{\Delta r^2}$ for the different $r_0$ cases, normalized according to the left-hand side of Eq.~\eqref{eq:ballistic_expanded}, and compare with the right-hand side shown as a dashed black line. For short times, $\tau \ll \tau_0$, the figure shows rather good agreement between the empirical data and Eq.~\eqref{eq:ballistic_expanded}, especially in the downward trend of the lines, which means that the above arguments of Ref.~\cite{Bitane2012} agree with our data. Furthermore, the negative term in Eq.~\eqref{eq:ballistic_expanded} is responsible for the deviations from a purely quadratic growth as time progresses. Therefore, defining the end of the ballistic regime as a deviation of $\av{\Delta r^2}$ by, say, 10\% from $\av{v_{||,0}} \tau^2$, we obtain that the ballistic regime ends at $\tau=0.2\,\tau_0$. Indeed, our data from different $r_0$ agrees with Eq.~\eqref{eq:ballistic_expanded} (dashed line) roughly up to $\tau\approx 0.2\tau_0$.

\begin{figure}[t!]
	\centering
	\includegraphics[height=5.5cm]{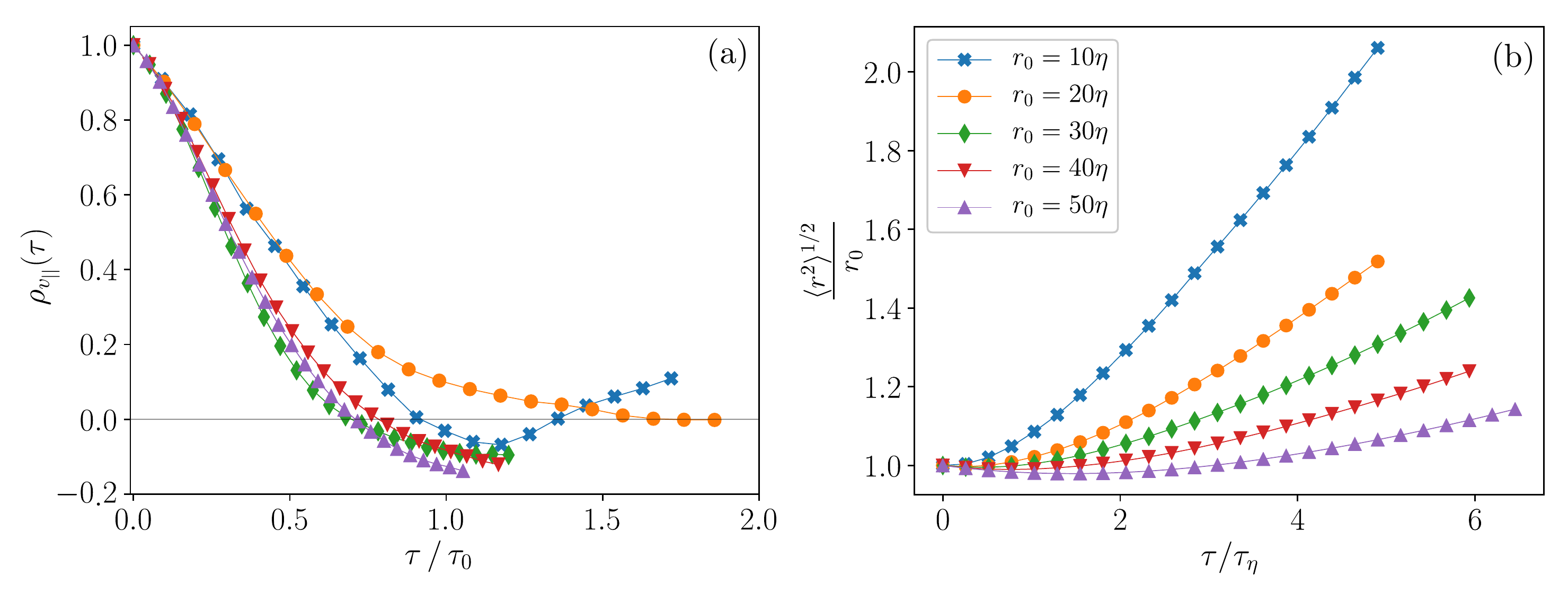}
	\caption{(a) Autocorrelation functions of the rate of separation $v_{||}$, plotted against time normalized with $\tau_0$. (b) root mean squared separation distance between trajectories, normalized with the initial separation, and plotted against time lag normalized with the Kolmogorov timescale. \label{fig:PD_b3_correlations}}
\end{figure}

Once the ballistic regime is terminated, the separation velocities of particles have deviated significantly from their initial values. This can be shown by examining the autocorrelation of the separation velocity, defined here as
\begin{equation}
\rho_{v_{||}}(\tau) = \frac{\av{v_{||}(\tau) \, \, v_{||}(0)}}{ \sqrt{\av{ v_{||}^2(0)} \, \, \av{v_{||}^2(\tau)}}} \,\, ,
\label{eq:sep_velocity_correlation}
\end{equation}
which is shown in Fig.~\ref{fig:PD_b3_correlations}. The autocorrelation $\rho_{v_{||}}$ is plotted against $\frac{\tau}{\tau_0}$ for the 5 cases of pairs with different $r_0$. The graph shows that the correlation of the separation velocity drops to roughly zero once the time lag is $\tau\sim\tau_0$ (with some weak dependence on $r_0$). This confirms that at $\tau \gtrsim \tau_0$ the particles' separation was with velocities that are very different from their initial value. We shall call this regime of the pair dispersion \emph{inertial}.

\subsection{Scale-dependence growth rates and approach to Richardson's 4/3 law}\label{sec:4/3_law}

Richardson's classical theory~\cite{Richardson1926} describes pair dispersion as an accelerating process in which the rate of separation increases with $r$. Accordingly, as $r$ increases with time, we should observe that the separation rate grows as well. Nevertheless, as discussed in the introduction, this is often challenging to observe in experiments and simulations due to the need for sufficiently long observation times such that $r$ increases significantly. To confirm whether such conditions occurred in our experiment we show in Fig.~\ref{fig:PD_b3_correlations}(b) the normalized separation, $r/r_0$, for five groups of pairs with different initial separations. For the group with the smallest initial separation available, $r_0=10\eta$, the typical separation increased by a factor of two by the end of our measurement range; on the other hand, the increase in separation was much lower for the other groups with larger $r_0$. This issue, however, is typical in experiments and simulations~\cite{Ott2000, Biferale2005, Liot2019, Elsinga2021, Tan2022}, because the separation timescale, $\tau_0 \sim r_0^{2/3}$, is smaller for smaller $r_0$. For this reason, if an explosive pair dispersion regime occurred in our experiment, it should be most evident in statistics of particles with small initial separations. Indeed, in Fig.~\ref{fig:PD_b3_longtime} (a), an increase of the local slope for the $r_0=10\eta$ group can be seen at longer times ($\tau \ge \tau_0$), indicating an increase in local scaling exponent. In these cases, the pair dispersion is super-diffusive, i.e. $\av{\Delta r^2} \propto \tau^\beta$ with $\beta>2$.

With the observations obtained thus far, let us lay out the reasoning to explain the behavior of pair dispersion in Fig.~\ref{fig:PD_b3_longtime} at longer times, namely after the Ballistic regime is over. First, for all cases of $r_0$ the separation velocity became decorrelated in our measurement (i.e. Fig.~\ref{fig:PD_b3_correlations}(a)), so that a so-called inertial regime was reached. Second, the increase of $r$ relative to its initial value depended on $r_0$, as shown in Fig~\ref{fig:PD_b3_correlations}(b). Third, in Richardson's theory pair dispersion is described as a diffusive process where the diffusivity depends on $r$ (i.e. $K_r(r)$). In the present case, pairs with small $r_0$ had multiplied their separation distance within the range of our measurement and therefore had increased their rate of separation, leading to the super-diffusive scaling $\beta>2$.
In contrast, for the largest $r_0$ case there was $\frac{\av{r^2}^{1/2}}{r_0}\approx 1$ for the entire range of our measurements. Therefore, in Richardson's diffusive framework, these pairs had an almost constant diffusivity, so the scaling exponent was $\beta \approx 1$, as expected in a diffusive process with constant diffusivity.

\begin{figure}[t!]
	\centering
	\includegraphics[width=\textwidth]{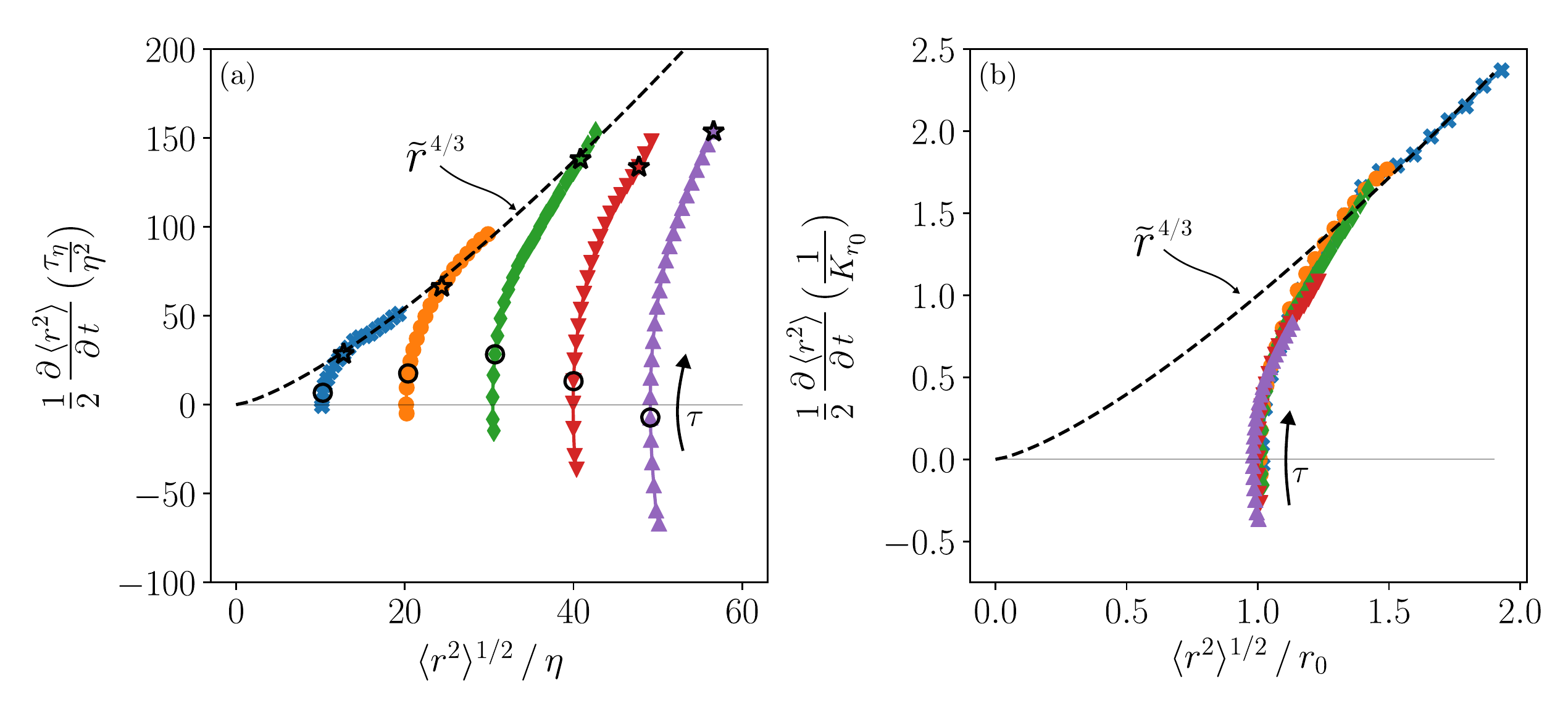}
	\caption{The diffusivity of pair dispersion defined as in Eq.~\eqref{eq:K_r} for various $r_0$ cases, plotted against the RMS of the separation distance; (a) data normalized by dissipation scales and (b) by inertial range scaling. Dashed lines correspond to the 4/3s law scaling, Eq.~\eqref{eq:k_scaling}. \label{fig:PD_b3_diffusivity}}
\end{figure}

We support the above picture by estimating diffusivity directly. Following Batchelor~\cite{Batchelor1952}, we define
\begin{equation}
K_r(r) \equiv \frac{1}{2} \frac{\partial \av{r^2}}{\partial t} \,\, .
\label{eq:K_r}
\end{equation}
In Fig.~\ref{fig:PD_b3_diffusivity}(a) the diffusivity is plotted against the RMS $\tilde{r} \equiv \av{r^2}^{1/2}$, where the axes are normalized using the dissipation scales, $\eta$ and $\tau_\eta$. An arrow on the figure indicates the direction in which the time grows. For all cases shown, $K_r$ is initially negative, which, as was discussed by \cite{Pumir2001}, is a consequence of the negative skewness of the Eulerian velocity differences, aka the 4/5 law~\cite{Monin1972}. With increasing time $K_r$ grows. At small times $K_r$ grows in line with the Ballistic regime, the end of which, at $\tau=0.2\tau_0$, is indicated as circles on the figure. Later on, at $\tau\approx\tau_0$, the separation velocity became decorrelated which is followed by the inertial regime. The time $\tau=\tau_0$ is indicated by stars on the figure. The range of our measurement extended to this regime ($\tau\gtrsim\tau_0$) for the cases of small $r_0$.

In the inertial regime, most pronounced for the small $r_0$ cases, the curves tend to the right side, meaning that as $\tilde{r}$ increased, the diffusivity increased as well. This is the main mechanism leading to the accelerating pair dispersion picture. Furthermore, the Richardson 4/3 law (as interpreted for the averaged separation by Batchelor~\cite{Batchelor1952}), according to eq.~\eqref{eq:k_scaling} predicts that $K_r \propto \tilde{r}^{\,4/3}$.
The prediction eq.~\eqref{eq:k_scaling} is plotted in the figure as a dashed black line. The data for the two smallest $r_0$ cases leans towards the dashed line for $\tau>\tau_0$. This observation agrees with previous results that showed a super-diffusive, $t^3$ scaling range for small initial separations in homogeneous isotropic turbulence~\cite{Elsinga2021, Tan2022}.

The Kolmogorov scaling for the diffusivity in the inertial range is $K_r \sim (\tilde{r}^{\,4} \, \epsilon)^{1/3}$. Therefore, we annotate $K_{r_0} \equiv (r_0^{4} \, \epsilon)^{1/3}$. Accordingly, in Fig.~\ref{fig:PD_b3_diffusivity}(b) the same data is plotted where the axes are normalized using $r_0$ and $K_{r_0}$. The data from all $r_0$ cases is seen to collapse on a single curve under this normalization. The collapse of the data from all $r_0$ in Fig.~\ref{fig:PD_b3_diffusivity}(b) confirms the existence of a scale-dependent regime in our pair dispersion measurement. This is a key result of this work. Also, as before the data for the smallest $r_0$ case joins the $K_r = K_{r_0} \, (\tilde{r}^{\,4} \, \epsilon)^{1/3}$ line which corresponds to Richardson's 4/3 law, Eq.~\eqref{eq:k_scaling}. These observations extend the fundamental phenomenology of turbulent pair dispersion to small-scale separation in highly turbulent anisotropic flows.

\section{Discussion and Conclusions}

Our work presents the first empirical investigation of turbulent pair dispersion in a canopy flow. For the analysis, we have used an extensive dataset of Lagrangian trajectories measured in a wind-tunnel experiment using an extended 3D-PTV method~\cite{Shnapp2019}.

The first part of our analysis focused on the anisotropy of pair dispersion. 
Our investigation is focused on small scales, being of the order of separation between canopy obstacles. As opposed to the anisotropy of canopy flow in larger scales,  our measurements show that in the canopy, at the scales of separation between canopy obstacles, pair dispersion anisotropy is weak.
The disparity among the separation components reached only up to approximately 20\% of the total separation. This was the result of our measurements being focused on scales smaller than $L_\Gamma$ - the scale that determines the crossover between small-scale local isotropy and large-scale anisotropy~\cite{Celani2005}. In addition, the time range of our measurement was smaller than $\Gamma^{-1}$, so even though particles separate with time and the scales to which they are exposed grow, their separation was still insufficiently large for anisotropy to build up and become prominent.

The weak anisotropy that does exist was observed to grow with time and initial separation. Furthermore, the spanwise dispersion inside the canopy was observed to be the strongest contributor to anisotropy in our measurements, presumably due to flow channeling in between the roughness obstacles. Investigating the pair dispersion tensor in a framework analogous to the return to isotropy~\cite{Lumley1977, Stiperski2021} (Fig.~\ref{fig:anisotropy_triangle}), showed that the anisotropy trajectories of pair dispersion progress towards anisotropy on a path that approximates oblate spheroids. Interestingly, in a recent meta-analysis of the Reynolds stress tensors anisotropy trajectories across numerous surface layer flows, a similar trajectory along the oblate topology was observed~\cite{Stiperski2021}. This similarity might hint that theory developed in the framework of ``return to isotropy'' could provide helpful information on pair dispersion modeling in anisotropic flows; however, this is left for future work.

An important feature observed in our experiments was the smallness of turbulence scales compared to those of mean shear. Due to this feature, the locally isotropic turbulent fluctuations were sufficiently strong to drive the nearly isotropic pair dispersion. This highlights the importance of $L_\Gamma / \eta$ and $T_L\,\Gamma$ to understanding pair dispersion in anisotropic turbulent flows, as having high and low values respectively implies that the isotropic turbulence phenomenology is the appropriate tool for analysis.

The second part of our analysis focused on the next two regimes of pair dispersion using the conceptual framework of local isotropy. At short times, the particles follow ballistic trajectories with separation velocities dictated by the Eulerian structure-function, as observed in homogeneous turbulence~\cite{Bourgoin2006}. At later times, the separation velocity undergoes decorrelation and a so-called inertial regime ensues. The rate of separation was shown to depend on the scale, $r$. For pairs with small initial separations, the separation rate was seen to scale in accordance to Richardson's 4/3 law ($r_0=10\eta$), in agreement with previous observations in isotropic turbulence~\cite{Elsinga2021, Tan2022}. These observations confirm that the leading mechanisms that govern pair dispersion at small scales in canopy flows are inherently tied to the internal turbulence regulation where the kinetic energy grows with the scale.

Understanding pair dispersion is important for modeling dispersion in the environment, as it can be used to estimate the variance of passive scalar concentration fields from known sources. In particular, Cohen and Reynolds~\cite{Cohen2000}, proposed a Lagrangian stochastic modeling approach for pair dispersion in highly inhomogeneous turbulent flows such as canopy flows. Their approach is consistent with the necessary conditions introduced by Thompson \cite{Thomson1990}, and their model was shown to agree with the experimental results of a scalar released from a known source.
Their model also uses the hypothesis that in highly inhomogeneous flows the correlation function for relative velocities is "short". In our measurements of the correlation function (Fig.~\ref{fig:PD_b3_correlations}a), we show explicitly at what time lag values this hypothesis is valid. In particular, $\rho_{v_{||}}=0$ is reached for all tested initial separations at approximately $\tau\approx \tau_0$ (Fig.~\ref{fig:PD_b3_correlations}a). This suggests that the concentration variance can be estimated from Cohen and Reynolds dispersion model for times $\tau > \tau_{v_{||}} \approx \tau_b = \left( \frac{r_0^2}{\epsilon} \right)^{1/3}$.

Lastly, it is important to regard our interpretation of the results using Richardson's turbulent diffusivity framework. Indeed, turbulent pair dispersion is an intermittent process~\cite{Boffetta2002, Bitane2012, Scatamacchia2012, Shnapp_Liberzon:2018, Tan2022}, and this characteristic cannot be described using the diffusivity model since it precludes longer correlation times. Nevertheless, as this is the first empirical investigation of the topic in a canopy flow, it is of particular importance to lay out the leading mechanisms that govern the phenomenon, which include the growth of the kinetic energy with the scale. Since the diffusivity framework lends itself somewhat naturally to the description of this phenomenology, it is the one that was chosen. With this, the treatment of more refined characteristics of pair dispersion at small scales is left for future research. To that should be added the treatment of pair dispersion over larger scales, which could not have been examined using the presently available dataset.

\section*{Acknowledgments}

We are grateful to Meni Konn, Sabrina Kalenko, Gregory Gulitski, Valery Babin, Amos Shick and Mordechai Hotovely, for their assistance in the wind tunnel experiment. This study is supported by the PAZY grants 2403170 and 1372020.

\appendix

\section{Subvolume flow parameters}\label{app1}

The following Tab.~\ref{tab1} and \ref{tab2} present the flow parameters used in the calculations for each of the subvolumes. Detailed calculations of these properties are given in Ref.~\cite{Shnapp2020}. 

\begin{table}[ht!]
	\scriptsize
	\centering
	\begin{tabularx}{.57\textwidth}{ l c c c c c c c }
		\hline\hline
		$sv$ & $\tilde{u}$ [m/s] & $\epsilon$ [W/kg] & $\eta$ [mm] & $\tau_\eta$ [s] & $\lambda$ [mm] & $Re_\lambda$ & $H/\eta$ \\ \hline
		$a1$ & 0.42 & 0.201 & 0.36 & 0.009 & 14.14 &  398 & 277 \\
		$a2$ & 0.45 & 0.256 & 0.34 & 0.008 & 13.25 &  394 & 295 \\
		$a3$ & 0.49 & 0.304 & 0.32 & 0.007 & 13.23 &  428 & 308 \\
		$a4$ & 0.51 & 0.244 & 0.34 & 0.008 & 15.35 &  517 & 291 \\
		$a5$ & 0.62 & 0.239 & 0.34 & 0.008 & 19.01 &  784 & 290 \\
		$b1$ & 0.36 & 0.123 & 0.41 & 0.011 & 15.41 &  369 & 245 \\
		$b2$ & 0.42 & 0.193 & 0.36 & 0.009 & 14.18 &  392 & 275 \\
		$b3$ & 0.47 & 0.250 & 0.34 & 0.008 & 14.09 &  440 & 293 \\
		$b4$ & 0.50 & 0.233 & 0.35 & 0.008 & 15.63 &  523 & 288 \\
		$b5$ & 0.65 & 0.305 & 0.32 & 0.007 & 17.63 &  762 & 308 \\
		$c1$ & 0.42 & 0.248 & 0.34 & 0.008 & 12.59 &  350 & 292 \\
		$c2$ & 0.43 & 0.210 & 0.36 & 0.008 & 13.97 &  397 & 280 \\
		$c3$ & 0.46 & 0.286 & 0.33 & 0.007 & 12.82 &  390 & 303 \\
		$c4$ & 0.52 & 0.231 & 0.35 & 0.008 & 16.13 &  555 & 287 \\
		$c5$ & 0.62 & 0.245 & 0.34 & 0.008 & 18.66 &  766 & 291 \\
		$d1$ & 0.40 & 0.175 & 0.37 & 0.009 & 14.18 &  373 & 268 \\
		$d2$ & 0.40 & 0.175 & 0.37 & 0.009 & 14.32 &  380 & 268 \\
		$d3$ & 0.47 & 0.229 & 0.35 & 0.008 & 14.61 &  454 & 287 \\
		$d4$ & 0.50 & 0.218 & 0.35 & 0.008 & 15.99 &  530 & 283 \\
		$d5$ & 0.66 & 0.377 & 0.31 & 0.006 & 16.10 &  707 & 325 \\
		\hline\hline
	\end{tabularx}
	\caption{Turbulence parameters for each sub-volume for the $Re_\infty = 16\times 10^3$ case. \label{tab1}}
\end{table}

\begin{table}[ht!]
	\scriptsize
	\centering
	\begin{tabularx}{.57\textwidth}{ l c c c c c c c }
		\hline\hline
		$sv$ & $\tilde{u}$ [m/s] & $\epsilon$ [W/kg] & $\eta$ [mm] & $\tau_\eta$ [s] & $\lambda$ [mm] & $Re_\lambda$ & $H/\eta$ \\
		\hline
		$a1$ & 0.53 & 0.422 & 0.30 & 0.006 & 12.16 &  426 & 334 \\
		$a2$ & 0.54 & 0.551 & 0.28 & 0.005 & 11.01 &  399 & 357 \\
		$a3$ & 0.60 & 0.703 & 0.26 & 0.005 & 10.68 &  424 & 379 \\
		$a4$ & 0.64 & 0.611 & 0.27 & 0.005 & 12.19 &  516 & 366 \\
		$a5$ & 0.83 & 0.669 & 0.27 & 0.005 & 15.19 &  839 & 375 \\
		$b1$ & 0.47 & 0.257 & 0.34 & 0.008 & 13.80 &  429 & 295 \\
		$b2$ & 0.50 & 0.352 & 0.31 & 0.007 & 12.74 &  427 & 319 \\
		$b3$ & 0.60 & 0.497 & 0.29 & 0.005 & 12.83 &  516 & 348 \\
		$b4$ & 0.64 & 0.487 & 0.29 & 0.006 & 13.79 &  589 & 346 \\
		$b5$ & 0.75 & 0.544 & 0.28 & 0.005 & 15.17 &  754 & 356 \\
		$c1$ & 0.52 & 0.490 & 0.29 & 0.006 & 11.05 &  379 & 347 \\
		$c2$ & 0.53 & 0.412 & 0.30 & 0.006 & 12.46 &  442 & 332 \\
		$c3$ & 0.60 & 0.587 & 0.28 & 0.005 & 11.68 &  464 & 363 \\
		$c4$ & 0.64 & 0.565 & 0.28 & 0.005 & 12.79 &  546 & 359 \\
		$c5$ & 0.81 & 0.706 & 0.26 & 0.005 & 14.49 &  783 & 380 \\
		$d1$ & 0.53 & 0.371 & 0.31 & 0.006 & 13.03 &  459 & 323 \\
		$d2$ & 0.52 & 0.327 & 0.32 & 0.007 & 13.72 &  478 & 313 \\
		$d3$ & 0.59 & 0.524 & 0.28 & 0.005 & 12.28 &  485 & 353 \\
		$d4$ & 0.63 & 0.528 & 0.28 & 0.005 & 12.94 &  540 & 353 \\
		$d5$ & 0.88 & 0.876 & 0.25 & 0.004 & 14.13 &  830 & 401 \\
		\hline\hline
	\end{tabularx}
	\caption{Turbulence parameters for each sub-volume for the $Re_\infty = 26\times 10^3$ case.  \label{tab2}} 
\end{table}

\pagebreak

\bibliography{bibliography}
\bibliographystyle{plain}

\end{document}